# Investigating Protected Health Information Leakage from Android Medical Applications

George Grispos[1], Talon Flynn[1], William Bradley Glisson[2], and Kim-Kwang Raymond Choo[3]

[1]University of Nebraska-Omaha, Omaha NE 68182, USA
[2]Sam Houston State University, Huntsville TX 77340, USA
[3]University of Texas at San Antonio, San Antonio TX 78249, USA

ggrispos@unomaha.edu; tflynn@unomaha.edu; glisson@shsu.edu; raymond.choo@fulbrightmail.org

**Abstract.** As smartphones and smartphone applications are widely used in a healthcare context (e.g., remote healthcare), these devices and applications may need to comply with the Health Insurance Portability and Accountability Act (HIPAA) of 1996. In other words, adequate safeguards to protect the user's sensitive information (e.g., personally identifiable information and/or medical history) are required to be enforced on such devices and applications. In this study, we forensically focus on the potential of recovering residual data from Android medical applications, with the objective of providing an initial risk assessment of such applications. Our findings (e.g., documentation of the artifacts) also contribute to a better understanding of the types and location of evidential artifacts that can, potentially, be recovered from these applications in a digital forensic investigation.

## 1    Introduction

Smartphone devices are becoming increasingly pervasive in today's medical environments. According to a survey conducted in the United States (US), just over 58% of surveyed individuals have downloaded a health-related mobile application on their smartphone [1]. Similarly, industrial surveys have estimated that more than 50% of physicians in the US encourage their patients to use smartphone medical applications [2, 3], particularly during medical emergencies (e.g., due to COVID-19 lockdowns [4, 5]). There is even evidence of increased usage of medical smartphone applications among healthcare professionals [6]. These smartphone applications can be used for a variety of different purposes, including disease self-management, remote monitoring of patients, as well as collecting and integrating patient data into Electronic Health Records (EHRs). However, the information collected and stored by these smartphone applications can also make them an attractive proposition for cybercriminals.

The 2020 Healthcare Information and Management Systems Society survey, for example, reports that 70% of healthcare organizations experienced a "significant security incident" in the past twelve months [7]. Further complicating matters, the US Food and Drug Administration (FDA) and the US

Government Accountability Office (GAO) have both warned that Internet-connected medical devices and their applications are likely to be susceptible to cyberattacks [8, 9]. These concerns were validated by the 2017 attacks on British hospitals when cybercriminals exploited unpatched and vulnerable Internet systems [10].

Hence, the security and privacy of patient information is an ongoing concern for the medical community [11, 12]. In 1996, the Health Insurance Portability and Accountability Act (HIPAA) became law in the US [13]. According to HIPAA, medical providers are required to implement administrative, technical, and physical safeguards in order to protect electronic Patient Health Information (ePHI) [14, 15]. More specifically, medical providers are expected to guarantee the confidentiality, integrity, and availability of ePHI, in order to protect this information. Patient and therapy data collected and stored by medical smartphone applications is likely to be considered as ePHI and, medical smartphone applications are likely to require additional security and privacy measures, as compared to typical smartphone applications.

Similar to other industries (e.g., banking, finance, and governments), the number of medical and healthcare applications that provide an Internet interface for medical devices is increasing [12, 16]. A number of these devices, and their accompanying smartphone applications, are FDA-approved for use in medical settings [17, 18]. However, mobile forensic researchers have demonstrated that many smartphone applications produce residual data, which introduces a variety of security and privacy challenges [19-21]. It has also been demonstrated that data stored within certain smartphone applications can be used by malicious actors to develop behavior profiles of the device user [22].

The increased usage of medical smartphone applications, in conjunction with continued healthcare system attacks, prompts the idea that these applications present security and privacy risks. It is also realistic to speculate that the residual data produced by medical smartphone applications is protected ePHI. This idea prompted the hypothesis that Android medical applications, which have interacted with medical devices, are putting user data at risk from a HIPAA security and privacy perspective. The contribution of this paper is twofold. Specifically, our findings support the idea that Android smartphone medical applications, which interact with medical devices, collect and store residual data that potentially violate HIPAA Security and Privacy regulations. The findings also contribute to the documentation of forensic artifacts recovered from specific Android medical smartphone applications.

The remainder of this paper is structured as follows. Section two presents related work, while section three describes the experimental design used to investigate the hypothesis. Section four presents the experiment results and an analysis of the findings, and section five concludes the paper and presents avenues for future work.

## 2    Related Literature

The widespread adoption of different consumer and operational technologies (e.g., mobile devices, and medical Internet of Things (MIoT) devices) into the healthcare domain has introduced a variety of security challenges and concerns [23]. Malasri and Wang [24] scrutinized implantable medical devices and describe how these type of devices are susceptible to various type of security attacks, which includes tracking patient information from the devices. Li et al. [25] reported that patient and device information is often transmitted in plaintext by medical devices, allowing an attacker to potentially recover device passwords and dosage information. Glisson et al. [26] explain that hospital teaching environments are also vulnerable to cyber-attacks by demonstrating brute force and denial-of-service attacks against a medical mannequin

The past few years has seen the emergence of MIoT devices [27]. However, data collected and transmitted by such devices (e.g., patient and therapy details) could potentially be targeted by malicious actors [12, 28]. Securing such data is particularly difficult when MIoT devices are deployed in environments that make it challenging to protect the underlying network, e.g., public Wi-Fi hotspots [27]. Data collected and compiled from different devices and sources could be cross-linked and subsequently be used to draw conclusions about a patient [28]. Hence, researchers have attempted to identify the challenges associated with the deployment of MIoT devices in settings such as hospitals and homes [29-31]. For example, Classen et al. [29] used a variety of techniques, including protocol analysis, software decompiling, and reverse engineering to extract private information from Fitbit fitness trackers. Similarly, Fereidooni et al. [30] reported that many of the fitness trackers included in their study did not include data integrity checks and implemented weak digital signatures. Wood et al. [31] demonstrated how an attacker can intercept unencrypted MIoT data transmissions, and use such information to profile the device owner.

Recent research also focuses on the integration of smartphone applications in the medical community. Miller et al. [32] scrutinized three telehealth smartphone applications and reported that two out of the three evaluated applications store information in plaintext. Similarly, Kharrazi et al. [33] reported that seven out of the nineteen mHealth applications evaluated in their study did not implement basic security features including password protection and user authentication mechanisms. Azfar et al. [34] put forth a forensic taxonomy which is based on a study of forty Android mHealth applications. This taxonomy consists of a variety of artifacts including databases, user credentials, and user activities.

Smartphone devices enhance the capabilities of traditional mobile phones and as a result, can provide forensic investigators with a wealth of potential evidence including web-browsing history, GPS coordinates, and third-party application-related data [35-37]. As a result, researchers have investigated how smartphone devices have been used to interact with social networks, cloud storage providers, and instant messaging services. Levinson et al. [38] demonstrated how it was possible to recover forensic artifacts from an iPhone device including device user information, GPS location information, and audio/video files. Al-Mutawa et al. [39] focused their research efforts on examining the forensic artifacts generated by smartphone social network applications and reported that Android devices produce detailed artifacts that describe social network activities. Anglano [40] investigated WhatsApp and reported that information recovered from its database artifacts can be used to reconstruct a timeline of the messages that have been exchanged between respondents. Grispos et al. [41, 42] analyzed smartphone cloud storage applications including Dropbox and Box, and reported data and metadata generated by these applications along with their storage paths. While previous research has investigated the recovery of digital artifacts from a variety of services and applications, minimal research has examined the artifacts that can be recovered from a medical device smartphone application, and how the storage of these artifacts, could potentially, violate HIPAA Security and Privacy regulations.

## 3 Experiment Design

A controlled experiment [43] was devised to investigate the extent to which residual data generated by Android medical applications, could potentially, violate Security and Privacy regulations within HIPAA. The experiment consists of five stages: 1) setup an Android smartphone and install the Android medical applications; 2) create test profiles on the medical applications and synchronize the associated medical devices with the applications; 3) use the medical applications and their associated medical devices; 4) process the smartphone device using a mobile phone forensic toolkit to create forensic images; 5) examine the resulting forensic image in order to recover artifacts related to the medical applications.

Six medical devices (hereafter referred to as the 'devices') from two manufacturers were evaluated in this experiment. These devices are described in Table 1 - Medical Devices. The devices were selected because they include an accompanying Android smartphone application and were available to the authors. The Android smartphone applications (hereafter referred to as the 'medical device applications') included in this experiment are iHealth MyVitals (version 3.7.1), iHealth Gluco-Smart (version 4.5.3), and Withings Health Mate (version 3.5.4).

Table 1. Medical Devices.

| Manufacturer | Device Name | Model | Mobile Application Name |
|---|---|---|---|
| iHealth | Core Scale | HS6 | iHealth MyVitals |
| iHealth | Feel Blood Pressure Monitor | BP5 | iHealth MyVitals |
| iHealth | Air Pulse Oximeter | PO3M | iHealth MyVitals |
| iHealth | Gluco-Monitoring System | BG5 | iHealth Gluco-Smart |
| Nokia | Cardio Scale | 03700546702341 | Health Mate |
| Nokia | BPM+ | 04719873310050 | Health Mate |

The smartphone used in this experiment is a Samsung Galaxy S4. This smartphone was selected for two reasons. First, the mobile forensic toolkit used in this experiment supports the smartphone and allows the extraction of a forensic image of the smartphone. Second, the smartphone executes Android as its operating system, which represents the most prevalent smartphone operating system at the time of the research [44]. While many smartphones could fulfill these criteria, the decision to use this specific device was based on availability and compatibility with the mobile forensic toolkit.

Two iterations of the experiment were executed. The first iteration involved the iHealth devices and applications, while the second iteration involved the Nokia devices along with their associated applications. The following steps were performed to prepare the smartphone, the devices, and medical device applications for the experiment.

1. The smartphone was restored to its factory settings. The smartphone was then powered-on, a Google account was created, and the initial setup process was completed.
2. The smartphone was connected to the Internet in order to download and install the medical device application via the Google Application store. The application was installed using the default setup options. A test profile was created along with a common password, for the experiment.
3. The medical device application was then used to 'pair' the medical device with the smartphone, using the instructions provided in the device user manuals. In the case of the iHealth Gluco-Monitoring System, additional steps were required before it can be paired with the smartphone. This device requires that a QR code is scanned, in addition to following the steps provided within its manual.
4. The medical devices and their accompanying medical device applications were then used by one of the authors for five days. In the case of the iHealth Gluco-Monitoring System, diluted sugared water was used instead of blood to measure glucose levels. In addition to the date and time of the recording, the following was also recorded for each medical device by observing the result on the device interface:
   - iHealth Core Scale: weight, body mass index, body fat, lean mass, body water, muscle mass, daily calorie intake, and bone mass.
   - iHealth Feel Blood Pressure Monitor: pulse, systolic, and diastolic values.
   - iHealth Air Pulse Oximeter: pulse, oxygen level, and perfusion index
   - iHealth Gluco-Monitoring System: blood sugar level.
   - Nokia Cardio Scale: weight, body fat, body water, pulse, bone mass, muscle mass, and body mass index.
   - Nokia BPM+ Device: pulse, systolic, and diastolic values.
   
   It must be noted that test information was used for the above measurements.
5. After the five-day period, a forensic image of the smartphone's internal memory was created using MSAB XRY (version 7.7). The instructions provided by XRY were completed and the smartphone's internal memory was read. The resulting forensic image was then saved to a desktop computer.
6. The forensic image, created in Step 5, was then examined using MSAB's accompanying analysis software, XAMN (version 3.2). This analysis involved locating files and artifacts related to the medical device application. Certain files and artifacts were exported from XAMN and examined further using FTK (version 6.3)

These steps were executed for each medical smartphone application on the smartphone. It must be noted that this research is limited in the following ways. The smartphone used in the experiment was purchased in the United States (US). Therefore, the smartphone contains software for US mobile phone carriers. Due to time constraints, the experiment was executed only once for each medical device and its accompanying smartphone application. Specific versions of the Android operating system and medical device smartphone applications were used in this experiment. Due to tool limitations, the version of the Android operating system was limited to Android version 5.0, in order to execute a Physical Extraction for the Android device. Furthermore, the experiment was limited to the application versions available at the time of the experiment. Finally, for this investigation, this experiment implemented replicated test data versus using real-world patient data.

## 4 Results and Analysis

The analysis of the files and artifacts generated by the medical device applications revealed information about the test patient and their use of the medical devices evaluated in the experiment. Within the Android operating system, application data and metadata are stored under the path `/data/data`. Applications installed on an Android device create folders under this location. The following subsections summarize the data retrieved from this analysis. These results only include the artifacts readily identifiable as associated with the medical device applications of interest and do not include all artifacts retrieved from the smartphone.

### 4.1 iHealth Applications

The iHealth MyVitals application generates a folder called `iHealthMyVitals.V2`. Artifacts related to the iHealth Core scale, the Feel Blood pressure monitor, and the Air pulse oximeter were recovered from this folder location. Within this folder is a subfolder called `Databases`, which contains a SQLite database called `androidNin.db`. Plaintext data parameters related to the iHealth devices can be retrieved from tables within this database. The first table of interest is called `TB_BPResult`. This table includes data related to the blood pressure monitor, including timestamps of the user's measurements, the systolic and diastolic values, the user's pulse rate, the device identifier, any text notes the user has entered as part of a reading, as well as the name of the user account associated with a particular reading.

Data parameters concerning the pulse oximeter can be retrieved from the database table called `TB_SPo2Result`. The data retrieved from this table (Fig. 1) includes the user's pulse rate ('PR'), perfusion index ('PI') and oxygen level ('Result'), along with the username ('HealthID') and timestamp ('MeasureTime') of each reading. Two tables called `TB_TemperatureHumidity` and `TB_WeightOnlineResult` contain data related to the Core scale. The `TB_TemperatureHumidity` table contains the humidity, temperature, and level of lighting as recorded by the scale. The `TB_WeightOnlineResult` table contains data related to the user's weight, body mass index value, body fat percentage, percentage of body water, muscle mass, daily calorie intake, and bone mass. The username and timestamp of each reading with the scale can also be recovered from the `TB_WeightOnlineResult` table.

| UsedUserid | PhoneDataID | iHealthID | MechineType | MechineDeviceID | MeasureTime | LastChangeTime | PhoneCreateTime | Result | PR | PI |
|---|---|---|---|---|---|---|---|---|---|---|
| Filter | Filter | Filter | Filter | Filter | Filter | Filter | Filter | | | Filter |
| 0 | 5CF821DE... | medicaldevices2018exper@gmail.com | PO3M | 5CF821DED2ED | 1530829549 | 1530829596 | 1530829549 | 97 | 89 | 9.69999... |
| 0 | 5CF821DE... | medicaldevices2018exper@gmail.com | PO3M | 5CF821DED2ED | 1530884863 | 1530884878 | 1530884863 | 96 | 77 | 8.30000... |
| 0 | 5CF821DE... | medicaldevices2018exper@gmail.com | PO3M | 5CF821DED2ED | 1531090846 | 1531090859 | 1531090846 | 97 | 90 | 4.90000... |
| 0 | 5CF821DE... | medicaldevices2018exper@gmail.com | PO3M | 5CF821DED2ED | 1531169685 | 1531169699 | 1531169685 | 96 | 80 | 12.3000... |
| 0 | 5CF821DE... | medicaldevices2018exper@gmail.com | PO3M | 5CF821DED2ED | 1531259730 | 1531259755 | 1531259730 | 97 | 90 | 3.0 |

**Fig. 1.** Retrieval of Information from TB_SPo2Result Table

In addition to recovering iHealth device readings, a separate database table called `TB_Userinfo` contains data associated with the device user. This data includes the user's name, date of birth, timezone location, and email address. The iHealth MyVitals application also stores user and device data within XML files. Within these files, the patient's email address, device names, and MAC network addresses can be recovered. However, an interesting observation is the recovery of the user's authentication credentials in plaintext from the XML file called `sp_user_region_host_info.xml`. Fig. 2 below, shows the recovery of the authentication access token and user password, stored in plaintext.

```xml
<map>
    <boolean name="medicaldevices2018exper@gmail.com_user_is_online" value="false"/>
    <string name="medicaldevices2018exper@gmail.com_user_refresh_token">
        D2PXXZMSfnfpbWDALTvmpUw0VRnZXDOdjicG9CT0QPuODtvy-BEeWr8wjr7coVdcAvge0t0-
        zTYf6ier3jyPQiUT5u7otC*4ZrrkVzYkx85LyoS9DtftXM-ig0*qcbcHx*RtLim-BlK7tZfzcoXtWg
    </string>
    <string name="medicaldevices2018exper@gmail.com_user_access_token">
        YgQLYRcdlyAWpL7cBiNLoORlyXd-uTznbuJaZRXiQgMb8EfRHdEF2q-hS0f2dQ0ryf*ubXJmrrUwG3RzY?
        sZEZV9f3ZoWYCglzSXbIgjPupID*X1eiJwi2JKVf7dw
    </string>
    <string name="medicaldevices2018exper@gmail.com_user_password">MedExp2018</string>
    <int name="medicaldevices2018exper@gmail.com_user_region_flag" value="1"/>
</map>
```

**Fig. 2.** Retrieval of Plaintext Password from iHealth Account

The iHealth Gluco-monitoring system creates a separate folder called `jiuan.an-droidBg.start` to store data related to this device. However, while various databases were recovered within this parent folder, all these databases appeared to be encrypted. Hence, it is not possible to recover device readings related to the Gluco-monitoring system from its smartphone application. Breaking the encryption of the databases is considered out of scope for this research. However, user data can be recovered from the XML file called `user_info.xml`. This user data includes the patient's username, along with the device identifier that interacts with the smartphone application.

### 4.2 Nokia Application

Concerning the Health Mate application, data can be retrieved from a parent folder called `com.with-ings.wiscale2`. User and device data are primarily stored within three tables in a database called `withings-WiScale.db`. The first table of interest is called `devices` and contains data related to the Nokia Cardio Scale and BPM+ devices. Information that can be recovered from this table includes the device's MAC address, timestamps related to when the device was last used, the type and model of the devices, as well as the battery level at the time of the last recording for the particular device. Fig. 3 shows an example of information that can be recovered from the `devices` table.

| id | associationDate | lastUseDate | modifiedDate | macAddress | firmware | timezone | battery | type | model |
|---|---|---|---|---|---|---|---|---|---|
| Filter | Filter | Filter | Filter | Filter | Filter | Filter | Filter | Filter | Filter |
| 5595648 | 1541806236000 | 1542127729662 | 1542127635000 | 00:24:e4:5a:ee:6c | 431 | NULL | 77 | 4 | 43 |
| 5402710 | 1541806407000 | 1542070868000 | 1542093243000 | 00:24:e4:57:12:c4 | 1751 | America/Chicago | 78 | 1 | 6 |

**Fig. 3.** Nokia Device Information

A second table called `measure` contains the measurements undertaken using the Nokia scale and blood pressure devices. The patient's weight, body fat, body water, pulse, bone mass, muscle mass, and body mass index, along with pulse information, systolic, and diastolic values can be recovered from this table. Each measurement includes timestamp information describing when the measurement was undertaken. It must be noted that three values associated with the Nokia Scale were not found within the `measure` table. These values are two Body Mass Index calculations and one body fat reading.

Finally, patient-related data can be recovered from a third table called `users`. Patient data retrieved from this table include the patient's name, gender, birthday, along with the email address used during the initial application registration.

### 4.3 Analysis

After examining the device manufacturer's websites, all the applications evaluated in this experiment are purported to comply with HIPAA legislation. HIPAA's Privacy Rule states that "individually identifiable health information held or transmitted by a covered entity or its business associate, in any form or media, whether electronic, paper, or oral" must be protected [45]. The Privacy Rule calls this information "Protected Health Information (PHI) and includes an individual's past, present or future physical or mental health condition; the provision of healthcare to an individual; the past, present or future payment for the provision of healthcare to an individual; and common identifiers such as the patient's name, address, date of birth and social security number" [45].

The results from the analysis of the smartphone application data suggest that two (iHealth MyVitals and Health Mate) out of the three applications are potentially putting a user's information at risk. Table 2 highlights the PHI recovered from the smartphone applications evaluated in this experiment. The table shows that the iHealth MyVitals and Health Mate applications disclosed plaintext information about the patient (name and birthday), as well as health conditions (heart rates, blood pressure readings, etc.). Hence, the findings suggest that two out of three applications are, potentially, putting patient and medical information at risk from a HIPPA Privacy Rule perspective.

Likewise, according to the HIPAA Security Rule, devices that store ePHI should consider utilizing encryption technologies of appropriate strength to protect sensitive information [46]. However, the analysis of the iHealth MyVitals and Health Mate applications revealed that these applications are storing patient and device information in plaintext. Furthermore, HIPAA also requires that safeguards are taken to protect passwords that can be used to access HIPAA-covered information [46]. However, the results from this experiment have shown that it is possible to recover the plaintext password from the MyVitals application. Hence, while all three applications claim to be HIPAA compliant, only the Android iHealth Gluco-Smart application appears to fully comply with the Security Rules requirements necessary for storing information, such as passwords, patient, and therapy details.

Table 2. HIPAA PHI Stored in Evaluated Applications.

| Manufacturer | iHealth MyVitals | Gluco-Smart | Health Mate |
|---|---|---|---|
| Individual's past, present or future physical or mental health condition | ✓ | X | ✓ |
| The provision of healthcare to an individual | ✓ | X | ✓ |
| Past, present, or future payment for the provision of healthcare to an individual | X | X | X |
| Patient's name | ✓ | ✓ | ✓ |
| Patient's address | ✓ | X | X |
| Patient's social security number | X | X | X |
| Patient's date of birth | ✓ | X | ✓ |

Key: ✓ = Recovered from Application; X = Not Recovered from Application

The examination of the above data supports the hypothesis that medical device applications, which interact with medical devices, violate the Security and Privacy rules within HIPAA. The statement holds for two out of the three applications evaluated in this research. The experimental findings reveal that patient-specific information can be retrieved from the applications including the patient's name, date of birth gender, as well as weight and height information. In addition to patient information, data related to the patient's use of the medical device can also be recovered from the medical device applications. However, the availability of this data is conditional on the specific smartphone application. For example, in terms of the iHealth MyVitals and Health Mate applications, this information included a patient's pulse rate, systolic and diastolic values, oxygen level, perfusion index, body mass index, daily calorie intake, bone mass, weight, body fat, lean mass, and body water level. Moreover, while minimal medical device usage data was recovered from the iHealth Gluco-Smart application, user details can be recovered from XML artifacts generated by the application.

## 5   Conclusions and Future Work

Smartphone devices are becoming increasingly pervasive in today's medical environments. As a result, there are security and privacy concerns related to the data stored and transmitted by these devices, when used in medical contexts. The findings from this initial research have demonstrated how certain smartphone applications, which interact with medical devices, violate the Security and Privacy rules within HIPAA. Information that can be recovered from these applications includes a patient's personal details and their usage of the specific medical device. This information could, in theory, be exploited by a cybercriminal who can gather intelligence about an individual's medical history, well-being, and potential ailments. Such information can then be used to develop medical profiles or combined with residual data from other smartphone applications to develop detailed patterns of user behavior.

There are several potential avenues for future research. The authors intend to expand this study to include an analysis of other medical devices and smartphone operating systems. This investigation investigates the applicability of the security and privacy concerns identified in this paper to other medical technical and environmental contexts. Research also needs to examine the degree to which data and metadata can be recovered from uninstalled medical device applications and the medical devices themselves. The insight gained from this avenue of research helps organizations to alleviate ePHI data leakage when devices are decommissioned and/or the smartphone applications are removed from a user's smartphone device. Tshe results of future research in this area potentially provide insight into the development of relevant security measures, mitigation solutions, and the identification of necessary healthcare policy components.

## 6   Acknowledgments

This research was financially supported by the Nebraska Research Initiative (NRI). The statements, opinions, and content included in this publication do not necessarily reflect the position or the policy of the NRI, and no official endorsement should be inferred.